\documentclass[conference,a4paper]{APSIPA2021}
\usepackage{multirow}
\usepackage{cite}
\usepackage{amsmath}
\usepackage{cite,balance}
\usepackage{multirow, multicol,graphicx}
\usepackage{amsmath}
\usepackage[psamsfonts]{amssymb}
\usepackage{amsxtra}
\usepackage{threeparttable}

\begin{document}

	\title{Processing Phoneme Specific Segments for Cleft Lip and Palate Speech Enhancement }

	\author{%
		\authorblockN{%
			Protima Nomo Sudro\authorrefmark{1}, Rohit Sinha\authorrefmark{1}, S. R. Mahadeva Prasanna\authorrefmark{2}
		}
		
		\authorblockA{%
			\authorrefmark{1}
			Indian Institute of Technology Guwahati, Guwahati, India\\
}
		
		\authorblockA{%
			\authorrefmark{2}
			Indian Institute of Technology Dharwad, Dharwad, India\\
			E-mail:\{protima, rsinha\}@iitg.ac.in, prasanna@iitdh.ac.in}
	}
	
\maketitle
\begin{abstract}
The cleft lip and palate (CLP) speech intelligibility is distorted due to the deformation in their articulatory system. For addressing the same, a few previous works perform phoneme specific modification in CLP speech. In CLP speech, both the articulation error and the nasalization distorts the intelligibility of a word. Consequently, modification of a specific phoneme may not always yield in enhanced entire word-level intelligibility. For such cases, it is important to identify and isolate the phoneme specific error based on the knowledge of acoustic events.  Accordingly, the phoneme specific error modification algorithms can be exploited for transforming the specified errors and enhance the word-level intelligibility. Motivated by that, in this work, we combine some of salient phoneme specific enhancement approaches and demonstrate their effectiveness in improving the word-level intelligibility of CLP speech. The enhanced speech  samples are evaluated using subjective and objective evaluation metrics.
\end{abstract}

\section{Introduction}
Speech intelligibility is important for communication either it is in human-human interaction mode or human-machine interaction mode~\cite{kain2007improving,rudzicz2013adjusting,nakamura2012speaking}. Due to articulatory impairment, the intelligibility of pathological speakers are degraded and it hinders them from communicating effectively as speakers without speech pathology do~\cite{whitehill2002assessing}. Hence, researchers study to improve the intelligibility of the pathological speech from signal processing point of view~\cite{lai2019multi, biadsy2019parrotron, fu2017joint,aihara2013individuality, xiao2019reconstruction, bi1997application}. In this work, cleft lip and palate (CLP) speech enhancement is addressed. 

The CLP is a birth disorder which affects the speech production system.  The speech disorders occur even after clinical intervention due to velopharyngeal dysfunction, oronasal fistula, and mislearning~\cite{kummer2013cleft,peterson2001cleft,henningsson2008universal}. The CLP speech distortions are categorized into hypernasality, hyponasality, articulation error and voice disorder~\cite{kummer2013cleft,scipioni2009intelligibility,maier2006intelligibility}. Among the many speech disorders associated with CLP speech, nasalization and articulation error are the primary factors that affect the speech intelligibility. Hypernasality corresponds to a resonance disorder, and the presence of nasal resonances during speech production has an excessively perceptible nasal quality~\cite{grunwell2001speech}. Mostly, the voiced sounds are nasalized, and the nasal consonants tend to replace the obstruents due to severe hypernasality. Misarticulations are produced either due to the structural or functional disorder or both. 

In the pathological speech enhancement literature, some of the reported works, focus on the segmentation of the disordered phoneme followed by modification of the same~\cite{vikram2016spectral,sudro2021modification,sudro2021event,rudzicz2013adjusting}. In several other studies, specific enhancement strategies are developed based on the disordered phenomena. In the aforementioned studies, phoneme specific modifications are attempted. These studies were effective for isolated phoneme analysis and modification. In contrast, considering a more realistic situation, the speech intelligibility and quality must be analyzed for words, and phrases. In clinical settings, when CLP speech is analyzed for enhancement tasks, the speech language pathologists (SLP), first work on isolated phonemes. Once a CLP speaker has mastered the correct phoneme production, the SLPs embed the phoneme in a word and analyze the speech intelligibility of an entire word. Further, this process is observed in a phrase and conversational speech. In a study, it is reported that a speaker may produce a phoneme correctly in isolation, but the same may be produced in error in a word-level and short phrases due to the influence of phonetic contexts~\cite{henningsson2008universal}. Hence, the SLPs evaluate and attempt to enhance the CLP speech by minimizing the influence of phonetic contexts. In the similar direction, the present work also focuses on the word-level intelligibility via combination of phoneme specific modification techniques reported in our earlier works~\cite{sudro2021modification,sudro2021event,sudro2020enhancement}. In those works, we have demonstrated the ability to modify the phoneme specific distortions for fricative /s/, stop consonant /k/, /t/, /T/ and vowels /a/, /i/, /u/. In the present work, the relevant phoneme specific enhancement techniques are combined to  improve the entire word-level intelligibility. 

\section{Transforming word-level speech intelligibility}
In this work, nonsensical word modification is attempted by combining the specific phoneme modification techniques  discussed in the our earlier works. Several words are formed by the possible combinations of the phonemes studied in the reported works. When a single transformation method is used to modify the entire word, it is observed that the speech sounds muffled. Thus, further processing is required to achieve a good quality enhanced speech. various issues exist in performing the entire word-level intelligibility because phonemes are either substituted by other phonemes or nasal consonants or gets nasalized. Different types of other misarticulation also affects the CLP speech intelligibility and quality. It is challenging to detect such misarticulations in an unsupervised method. Therefore, certain assumptions are made prior to the enhancement task.

In CLP speech, the production of obstruents are mainly affected due to the loss of adequate intra-oral pressure or mislearned compensatory articulation.  Due to the production error, important acoustic-phonetic cues related to obstruents, such as transient burst, frication noise, and formant dynamics in the adjacent sonorant's transition region are degraded. Therefore, it is expected that deviations in their productions, that are reflected on the acoustic signal may be correlated with the perceived CLP speech intelligibility. On the other hand, due to nasalization, the voiced sounds are mostly affected. In a word, when the obstruents are modified, distorted voiced sounds could still affect the overall intelligibility. This renders the modification of the voiced sounds as well. Hence, characterizing all the  deviations using relevant acoustic features and then modification of the same are expected to provide intelligible speech. The knowledge of deviated acoustic characteristics explored for the analysis of CLP speech in the previous works~\cite{sudro2020enhancement,sudro2021event,sudro2021modification} can be used to enhance the distorted word intelligibility. 

\subsection{Database description}
The database was created in collaboration with the speech language pathologists (SLPs) of All India Institute of Speech and Hearing (AIISH), Mysuru, India. The database consists of age and gender matched CLP and non-CLP speakers’ speech. The non-CLP speakers served as controls for the study. Both the CLP and non-CLP children are native Kannada speakers. Prior to the recording,
ethical consents were obtained from the parents/caregivers of each of the participants. An overview of the study is provided to the parents/caregivers. The study was conducted with clearance from the AIISH Bio-behavioral ethical committee. In this work, nonsensical words are used for analysis and modification. All the speech samples are recorded in a sound-proof room using a speech level meter (Bruel and Kj\ae r) at a sampling frequency of 48 kHz and 16-bit resolution~\cite{slm_recording}. The recorded speech samples are saved in a .WAV format. The microphone was placed at a distance of 15 cm from each speaker while recording. During recording, the instructor first uttered the target word, then the response of the children was recorded. Each of the expert SLP had an experience of around five years in the field of CLP speech evaluation. The speech samples are recorded for 2-3 sessions for each of the speaker. The total number of speakers, number of tokens, and assessment of the distorted samples for fricatives, stops, and vowels can be found in~\cite{sudro2021modification}, ~\cite{sudro2021event}, and~\cite{sudro2020enhancement} respectively.


\subsection {Modification of fricative-vowel-fricative-vowel words}
Considering the combination of misarticulated fricative /s/ and vowels /a/, /i/, and /u/ as /sasa/, /sisi/, and /susu/, the enhancement task must address all the distorted phonemes to improve the overall speech intelligibility and quality~\cite{kummer2013cleft,peterson2001cleft}. The enhancement techniques exploited in the previous works, such as spectral energy compression~\cite{sudro2021modification}, insertion~\cite{singh2016structure}, spectral conversion~\cite{stylianou1998continuous} and temporal processing~\cite{krishnamoorthy2011enhancement} are briefly discussed for word-level intelligibility. The experimental details of the aforementioned techniques can be found in~\cite{sudro2020enhancement,sudro2021event,sudro2021modification}.
  \begin{figure}[!h]
  	{\centerline
  		{\includegraphics[width=0.51\textwidth]{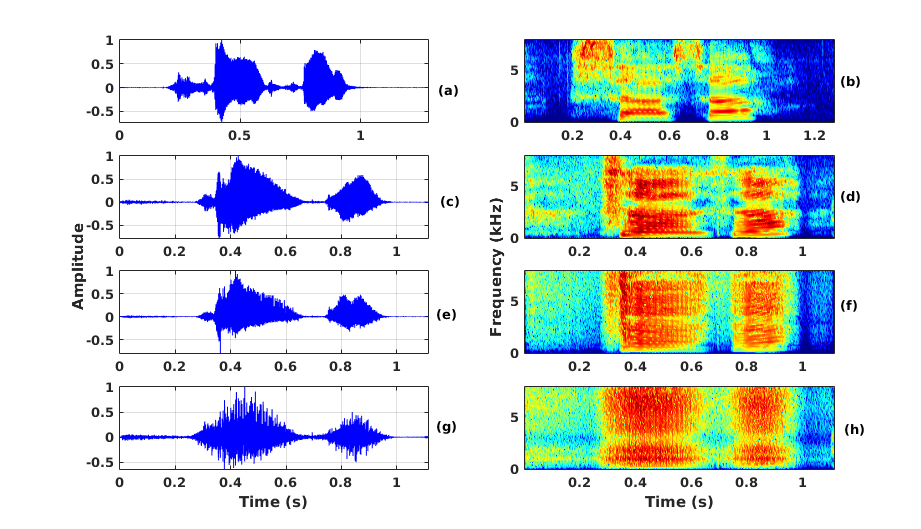} }
  		\caption{ Comparison of (a)-(b) non-CLP (healthy) /sasa/, (c)-(d) NAE distorted /sasa/,  (e)-(f) enhanced /sasa/ using GMM based spectral conversion method and (g)-(h) enhanced /sasa/ using NMF based spectral conversion method .}
  		\label{s_gmm_nmf}
  	}
  \end{figure}
  
If spectral compression technique is to be applied to modify the entire /sasa/ or /sisi/ or /susu/ word, then lower frequency region compression will yield enhanced fricatives but it will deteriorate the vowels at the same time because low-frequency energy is important for vowel perception~\cite{stevens2000acoustic,narayanan2000noise, shadle1985acoustics, shadle1996quantifying}. In certain cases, if vowels are nasalized, insertion method may be applicable but its perceptual quality will be far from the natural speech as it may not preserve the individuality information. Considering the temporal processing method described in~\cite{krishnamoorthy2011enhancement}, a fine weight function is applied around the glottal closure instants (GCI) locations to emphasize the significant GCI events and suppress any interfering signal components around it. If the fricative-vowel-fricative-vowel (FVFV) words  are processed using temporal processing method, it is speculated that it might not result in effective transformation because the excitation source for fricative is a noise source signal and with temporal processing important signal components might get suppressed. As each of the phonemes have different spectral and temporal phenomena, the distortions caused by the articulatory impairment affects each phoneme differently. Further, considering the spectral conversion method, mapping the distorted CLP spectra to a desired speech spectra is a good strategy towards attaining CLP speech enhancement. Hence, an attempt is made to enhance /FVFV/ words using Gaussian mixture modeling (GMM) based spectral conversion method~\cite{stylianou1998continuous}. 

The enhanced /sasa/ word using GMM based spectral conversion is depicted in Fig.~\ref{s_gmm_nmf} (e)-(f). After performing enhancement, it is observed that GMM based spectral conversion of hypernasal speech results in muffled speech. Further analysis of the modified speech signal showed that the spectrally modified speech  is observed to have low nasalization compared to original /FVFV/ word. In Fig.~\ref{s_gmm_nmf} (c)-(d) the fricatives are ambiguous and the overall speech quality is still degraded.

The speech degradation may be attributed to the oversmoothing effect. Because of the oversmoothing effect the formants of the vowels are observed to have larger bandwidth, smaller peak-to-valley ratio, and the fricative spectrum are observed to have deviant acoustic characteristics and spectral tilt. Several techniques are proposed in the literature to overcome the smoothening affect, where one of the approach corresponds to the parametric voice conversion (VC) method which is reported to alleviate the oversmoothing effect. Therefore, nonnegative matrix factorization (NMF) based speech enhancement is used to modify the /FVFV/ word and shown in Fig.~\ref{s_gmm_nmf} (g)-(h)~\cite{fu2017joint}. The modified /FVFV/ word shows some improvement compared to original /FVFV/ word but its acoustic characteristics are still far from that of the non-CLP /FVFV/ word shown in Fig.~\ref{s_gmm_nmf} (a)-(b). Although some improvements are observed as reduction in nasalization and spectral prominence in the high-frequency region for fricatives. However, further observation showed that low frequency energy in the fricatives persists and the formants of the vowels are not distinct. Thus, it projects the importance of processing different class of sound units separately. 


\subsection{Modification of consonant-vowel-consonant-vowel words}

The consonant-vowel-consonant-vowel (CVCV) words consisting of the stops /k/, /t/ and /T/ in vowel contexts /a/, /i/, and /u/ are analyzed similar to that of the /FVFV/ described in previous section. The possible combination of words are /kaka/, /kiki/, /kuku/, /tata/, /titi/, /tutu/, /TaTa/, /TiTi/, and /TuTu/. The spectral compression technique may not effectively transform the /CVCV/ word because the spectral prominence of consonants are not confined to a specific frequency band. For example, the velar stop /k/ is characterized by prominent spectral energy in the low-frequency region, whereas, the alveolar stop /t/ shows spectral prominence around mid-frequency region and retroflex /T/ shows spectral prominence above 2~kHz. Additionally, formants of the vowels are observed in the low-frequency region. Therefore, spectral energy compression in the low-frequency region will result in further deterioration of the /CVCV/ words. Further, temporal processing will result in vowel modification only and using insertion method artificially synthesized phoneme can be used for speech modification. However, insertion method does not preserve the individuality information. Hence, spectral compression, temporal processing and insertion method might not be effective in transforming the entire /CVCV/ word. 
  \begin{figure}[!h]
  	{\centering
  		{\includegraphics[width=0.52\textwidth]{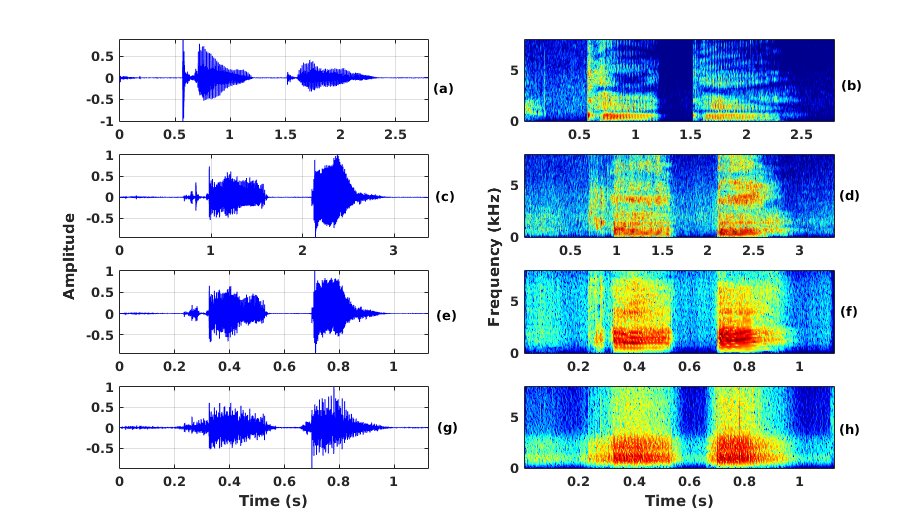} }
  		\caption{ Comparison of (a)-(b) non-CLP /kaka/, (c)-(d) misarticulated /kaka/, (e)-(f) enhanced /kaka/ using GMM based spectral conversion method and (g)-(h) enhanced /kaka/ using NMF based spectral conversion method.}
  		\label{c_gmm_nmf}
  	}
  \end{figure}
The consonants are very short duration phonemes with dynamic characteristics that represents different spectral and temporal importance~\cite{prathosh2014estimation,delattre1955acoustic}. The spectral prominence for consonants ranges from low-frequency region to high-frequency region and in vowels the lower frequency region are mostly considered to carry important perceptual information~\cite{singh2016structure}. Hence, the modification technique designed for a specific category of phoneme may or may not be effective in transforming the disordered nature of other phoneme. As stated in the previous section, mapping the disordered speech spectra into that of the non-CLP speech spectra may improve the speech intelligibility and quality. Hence, the modification of /CVCV/ word is also attempted using GMM and NMF based spectral conversion respectively.

For illustration, the transformed signals for the word /kaka/ is shown in Fig.~\ref{c_gmm_nmf}. From the figure, it is observed that the vowel formants and the consonants in Fig.~\ref{c_gmm_nmf} (e)-(f) and Fig.~\ref{c_gmm_nmf} (g)-(h) are not similar to non-CLP speech characteristics shown in Fig.~\ref{c_gmm_nmf} (a)-(b). Based on the above figures, it is observed that the enhanced speech signal does not show significant improvement relative to the original unprocessed signal. 
The reason may be attributed to the complex relationship of the misarticulations and nasality in CLP speech. Hypernasality and articulation error both show an impact in the same word reducing the speech intelligibility and quality.
In the above figures, for an illustration only /kaka/ is depicted. However, similar analysis are observed for other misarticulations in the stop /t/ and /T/, respectively. Both the NMF and GMM based spectral conversion had shown some improvement in the speech characteristics, however, they are not able to effectively enhanced the speech signal as desired. Therefore, this necessitates further analysis of the signal characteristics and then perform modification. Hence, phoneme specific enhancement can be attempted to observe the impact on the overall speech intelligibility and quality of a word.

\section{Experimental evaluation}

In this section, the impact of the combined phoneme specific modification techniques are analyzed. From Fig.~\ref{s_gmm_nmf} and Fig.~\ref{c_gmm_nmf}, it is observed that when entire word is modified, all the phoneme distortions (either fricative or consonant misarticulation or nasalization) were not reduced as desired.

\begin{table*}[!tbh]
	\centering	
	\caption{\footnotesize{P-STOI values for different combination of nonsensical words. F denote fricative /s/, C$_1$ denote consonant /k/, C$_2$ denote consonant /t/, C$_3$ denote consonant /T/, GS denote glottal stop substitution, PA denote palatalized articulation, PSNAE denote phoneme specific nasal air emission, and velar denote velar substitutions.} }
	\label{wrd_p_stoi}
	\scalebox{1.05}{
		\begin{tabular}{ccccccccccc}
			\hline \hline
			\multirow{2}{*}{ Words} &  & \multirow{2}{*}{Normal  Reference} &  & \multirow{2}{*}{CLP$_{\text{original}}$} &  & \multicolumn{5}{c}{CLP$_{\text{modified}}$}          \\ \cline{7-11} 
			&  &                                    &  &                                 &  & Obstruent &  & Vowel &  & Obstruent + vowel \\ \hline		
			/FVFV/ (GS)           &  & 0.88                   &  & 0.02           &  & 0.27                   &  & 0.23                     & & 0.46  \\
			/FVFV/ (PA)           &  & 0.88                   &  & 0.11           &  & 0.26                   &  & 0.22                     & & 0.44 \\
			/FVFV/ (PSNAE)        &  & 0.88                   &  & 0.12           &  & 0.35                   &  & 0.21                     & & 0.49  \\ \hline
			/C$_1$VC$_1$V/ (GS)   &  & 0.84                   &  & 0.04           &  & 0.11                   &  & 0.33                     & & 0.40 \\ \hline
			/C$_2$VC$_2$V/ (GS)  &  & 0.81                   &  & 0.12           &  & 0.14                   &  & 0.16                     & & 0.37 \\
			/C$_2$VC$_2$V/ (Velar)&  & 0.81                   &  & 0.13           &  & 0.14                   &  & 0.18                     & & 0.39 \\
			/C$_2$VC$_2$V/ (PA)  &  & 0.81                   &  & 0.13           &  & 0.14                   &  & 0.19                     & & 0.42 \\ \hline
			/C$_3$VC$_3$V/ (GS)  &  & 0.82                   &  & 0.15           &  & 0.17                   &  & 0.24                     & & 0.43 \\
			/C$_3$VC$_3$V/ (Velar) &  & 0.82                   &  & 0.18           &  & 0.19                   &  & 0.23                     & & 0.39 \\
			/C$_3$VC$_3$V/ (PA)    &  & 0.82                   &  & 0.16           &  & 0.18                   &  & 0.18                     & & 0.42  \\  \hline\hline
		\end{tabular}
	}
\end{table*} 

\begin{figure}[!tbh]
	{\centerline
		{\includegraphics[width=0.51\textwidth]{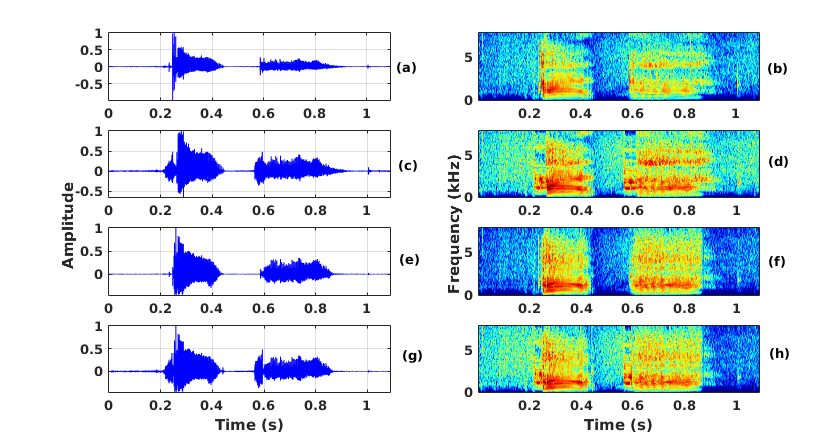} }
		\caption{ Illustration of waveform and spectrogram of: (a)-(b) clp /kaka/, (c)-(d) modified /k/, (e)-(f) modified /a/, and (g)-(h) modified /k/ and /a/.}
		\label{clp_kaka}
	}
\end{figure}

\begin{figure}[!tbh]
	{\centerline
		{\includegraphics[width=0.51\textwidth]{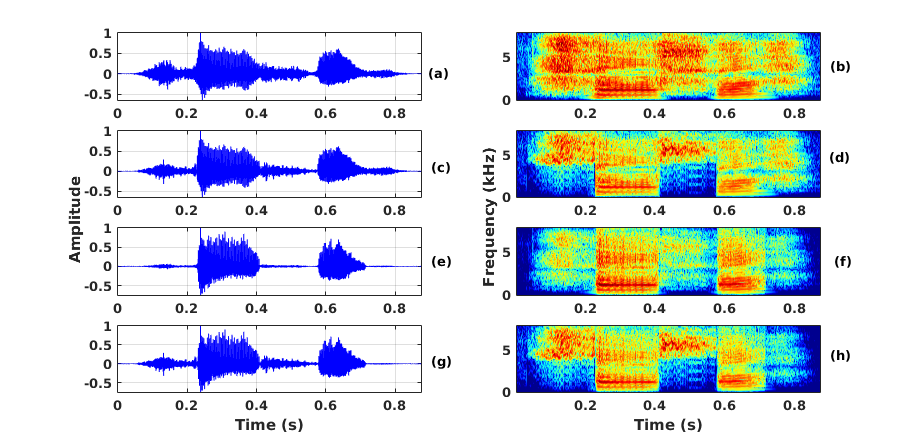} }
		\caption{ Illustration of waveform and spectrogram of: (a)-(b) clp /sasa/, (c)-(d) modified /s/, (e)-(f) modified /a/, and (g)-(h) modified /s/ and /a/.}
		\label{clp_sasa}
	}
\end{figure}

When each of the phoneme in a word are modified independently as shown in Fig.~\ref{clp_kaka} and Fig.~\ref{clp_sasa}, the impact of the speech distortions are reduced significantly. Due to severity, both articulation error and nasalization are observed in CLP speech. To have an enhancement system capable of performing the desired task, it is essential to improve the entire word intelligibility.

To analyze the effectiveness of the transformation method exploited in our earlier works, the evaluation is carried out for entire word modification and compared with isolated phoneme modifications.  At first the impact on word-level intelligibility is evaluated for the enhanced speech obtained by transforming only fricatives/consonants.

In the second step, the enhanced speech obtained using only vowel modification is evaluated. Finally, in the third step, the enhanced speech obtained by exploiting the consonant and vowel modification both are evaluated.

\subsection{Objective evaluation}
The modified CLP speech is assessed using  pathological short-time objective intelligibility (P-STOI) and  pathological extended short-time objective intelligibility (P-ESTOI) measures. The mathematical descriptions of P-STOI and P-ESTOI can be found in~\cite{janbakhshi2019pathological}.


The objective measures corresponding to P-STOI and P-ESTOI are noted in Table.~\ref{wrd_p_stoi} and Table.~\ref{wrd_p_estoi}, respectively. The P-STOI and P-ESTOI values are computed for time-aligned CLP speech signals and non-CLP (reference) signals. Before comparing the measures, word specific reference  templates are created from non-CLP speaker's speech. The reported values are averaged across all the listeners corresponding to the specific errors in all the vowel contexts. P-STOI values in Table~\ref{wrd_p_stoi} indicate that compared to original CLP speech misarticulation, modification of either of the error (obstruents misarticulation or vowel nasalization) results in improved intelligibility. However, from the P-STOI values, it is also observed that modification of both the distorted obstruents and vowels provide higher P-STOI values as compared to standalone modification. Similar observations are also noted for the P-ESTOI values tabulated in Table~\ref{wrd_p_estoi}.

\begin{table*}[!tbh]
	\centering
	\caption{\footnotesize{P-ESTOI values for different combination of nonsensical words. F denote fricative /s/, C$_1$ denote consonant /k/, C$_2$ denote consonant /t/, C$_3$ denote consonant /T/, GS denote glottal stop substitution, PA denote palatalized articulation, PSNAE denote phoneme specific nasal air emission, and velar denote velar substitutions.} }
	\label{wrd_p_estoi}
	\scalebox{1.05}{
		\begin{tabular}{ccccccccccc}
			\hline \hline
			\multirow{2}{*}{Words} &  & \multirow{2}{*}{Normal  Reference} &  & \multirow{2}{*}{CLP$_{\text{original}}$} &  & \multicolumn{5}{c}{CLP$_{\text{modified}}$}          \\ \cline{7-11} 
			&  &                                    &  &                                 &  & Obstruent &  & Vowel &  & Obstruent + vowel \\ \hline		
			
			/FVFV/ (GS)                &  & 0.25                     &  & 0.02             &  & 0.06                   &  & 0.13                      & & 0.18   \\
			/FVFV/ (PA)                &  & 0.25                     &  & 0.06             &  & 0.09                   &  & 0.16                      & & 0.22 \\
			/FVFV/ (PSNAE)             &  & 0.25                     &  & 0.01             &  & 0.02                   &  & 0.15                      & & 0.24  \\ \hline
			/C$_1$VC$_1$V/ (GS)                &  & 0.19                     &  & 0.03             &  & 0.07                   &  & 0.16                      & & 0.20  \\ \hline
			/C$_2$VC$_2$V/ (GS)                &  & 0.31                     &  & 0.09             &  & 0.15                   &  & 0.17                      & & 0.23  \\
			/C$_2$VC$_2$V/ (Velar)             &  & 0.31                     &  & 0.07             &  & 0.08                   &  & 0.10                      & & 0.16 \\
			/C$_2$VC$_2$V/ (PA)                &  & 0.31                     &  & 0.06             &  & 0.10                   &  & 0.17                      & & 0.23 \\ \hline
			/C$_3$VC$_3$V/ (GS)                &  & 0.29                     &  & 0.12             &  & 0.17                   &  & 0.20                      & & 0.22 \\
			/C$_3$VC$_3$V/ (Velar)             &  & 0.29                     &  & 0.15             &  & 0.15                   &  & 0.21                      & & 0.19  \\
			/C$_3$VC$_3$V/ (PA)                &  & 0.29                     &  & 0.01             &  & 0.07                   &  & 0.11                      & & 0.27  \\  \hline\hline
		\end{tabular}
	}
\end{table*}

For an illustration, the original and modified CLP words are also analyzed using automatic speech recognition (ASR) performance. As the speech data for this study are in the Kannada language, a Kannada ASR system is developed using KALDI speech recognition toolkit~\cite{povey2011kaldi}.  The ASR system performance for various phoneme modification categories is measured using phone error rate (PER) metric. The PER for the original and modified CLP speech is shown in Table~\ref{wrd_asr}. 
\begin{table}[!tbh]
	\centering
	\caption{\footnotesize{Phoneme error rate (\%) for various combination of nonsensical words. F corresponds to fricative /s/, C$_1$ corresponds to consonant /k/, C$_2$ corresponds to consonant /t/, C$_3$ corresponds to consonant /T/, GS corresponds to glottal stop substitution, PA corresponds to palatalized articulation, PSNAE corresponds to phoneme specific nasal air emission, and velar corresponds to velar substitutions.}}
	\label{wrd_asr}
	\scalebox{0.78}{
		\begin{tabular}{ccccccccc}
			\hline \hline
			\multirow{2}{*}{Words} &  &  \multirow{2}{*}{CLP$_{\text{original}}$} &  & \multicolumn{5}{c}{CLP$_{\text{modified}}$}          \\ \cline{5-9} 
			&  &                                                                    &  & Obstruent &  & Vowel &  & Obstruent + vowel \\ \hline
			/FVFV/ (GS)            &  & 68.85       &  & 56.08      &  & 58.15    &  & 54.53            \\
			/FVFV/ (PA)            &  & 59.23       &  & 38.80      &  & 36.83    &  & 31.58            \\
			/FVFV/ (PSNAE)         &  & 63.38       &  & 39.99      &  & 35.28    &  & 32.97            \\ \hline
			/C$_1$VC$_1$V/ (GS)    &  & 67.17       &  & 55.44      &  & 58.78    &  & 53.14           \\ \hline
			/C$_2$VC$_2$V/ (GS)    &  & 64.30       &  & 53.04      &  & 59.27    &  & 52.73          \\
			/C$_2$VC$_2$V/ (Velar) &  & 68.46       &  & 59.96      &  & 50.18    &  & 54.04          \\
			/C$_2$VC$_2$V/ (PA)    &  & 49.72       &  & 43.52      &  & 48.06    &  & 42.98         \\ \hline
			/C$_3$VC$_3$V/ (GS)    &  & 77.31       &  & 63.09      &  & 65.92    &  & 62.28          \\
			/C$_3$VC$_3$V/ (Velar) &  & 75.26       &  & 64.91      &  & 67.22    &  & 63.29         \\
			/C$_3$VC$_3$V/ (PA)    &  & 72.54       &  & 61.43      &  & 62.18    &  & 57.65         \\  \hline\hline
		\end{tabular}
	}
\end{table}
Compared to the original unprocessed CLP speech, the recognition performance of the modified speech is relatively higher. The modification of a specific phoneme also yield reduced PER value compared to that of the original distorted CLP speech. However, better performances are observed when both the phonemes in the word are modified. In certain instances, the PER of the combined modification is comparable to the standalone modification  of the phoneme. This implies that, in those utterances, the distortion caused by that specific phoneme is dominant.

\begin{table}[!tbh]
	\centering
	\caption{\footnotesize{MCD values for different combination of nonsensical words. F corresponds to fricative /s/, C$_1$ corresponds to consonant /k/, C$_2$ corresponds to consonant /t/, C$_3$ corresponds to consonant /T/, GS corresponds to glottal stop substitution, PA corresponds to palatalized articulation, PSNAE corresponds to phoneme specific nasal air emission, and velar corresponds to velar substitutions.}}
	\label{wrd_mcd}
	\scalebox{0.78}{
		\begin{tabular}{ccccccccc}
			\hline \hline
			\multirow{2}{*}{Words} &  &  \multirow{2}{*}{CLP$_{\text{original}}$} &  & \multicolumn{5}{c}{CLP$_{\text{modified}}$}          \\ \cline{5-9} 
			&  &                                                                    &  & Obstruent &  & Vowel &  & Obstruent + vowel \\ \hline
			/FVFV/ (GS)            &  & 13.00       &  & 11.20      &  & 11.80    &  & 11.90            \\
			/FVFV/ (PA)            &  & 12.94       &  & 11.31      &  & 12.23    &  & 12.29                        \\
			/FVFV/ (PSNAE)         &  & 13.00       &  & 11.50      &  & 12.18    &  & 12.26                        \\ \hline
			/C$_1$VC$_1$V/ (GS)    &  & 12.34     &  & 11.91        &  & 11.41    &  & 11.53                         \\ \hline
			/C$_2$VC$_2$V/ (GS)    &  & 12.73     &  & 10.81        &  & 11.91    &  & 11.98                        \\
			/C$_2$VC$_2$V/ (Velar) &  & 13.58     &  & 10.91        &  & 11.75    &  & 12.22                        \\
			/C$_2$VC$_2$V/ (PA)    &  & 13.58     &  & 10.58        &  & 11.90    &  & 12.13                       \\ \hline
			/C$_3$VC$_3$V/ (GS)    &  & 13.74     &  & 10.85        &  & 11.76    &  & 12.02                       \\
			/C$_3$VC$_3$V/ (Velar) &  & 13.80     &  & 10.67        &  & 11.87    &  & 12.27                      \\
			/C$_3$VC$_3$V/ (PA)    &  & 13.91     &  & 10.46        &  & 11.88    &  & 12.36                   \\  \hline\hline
		\end{tabular}
	}
\end{table}

Considering the MCD values shown in Table~\ref{wrd_mcd}, it is observed that the combined modification of the obstruent and vowels, results in lower MCD values relative to that of the original distorted CLP words. The MCD values reported in Table~\ref{wrd_mcd} are averaged across all the listeners corresponding to the specific errors in all the vowel contexts. The lower MCD values of the modified words indicate that the spectral difference between non-CLP (reference) and misarticulated words are reduced significantly for all the words evaluated in the study.

In certain cases, the objective intelligibility values of combined modifications are observed to be comparable with standalone modification. The probable reason may be attributed to the fact that either obstruent or vowel in the word is less distorted, hence resulting in comparable values.

\subsection{Subjective evaluation}
Listening experiment is also carried out to assess the word-level intelligibility by exploiting the phoneme specific modification techniques reported in the earlier works. The speech quality of the distorted CLP speech and the modified CLP speech is measured using a 5-point rating scale mean opinion score (MOS) where, 1 = bad, 2 = fair, 3 = good, 4 = very good, and 5 = excellent. 
\begin{table}[!tbh]
	\centering
	\caption{\footnotesize{MOS for different combination of nonsensical words. F corresponds to fricative /s/, C$_1$ corresponds to consonant /k/, C$_2$ corresponds to consonant /t/, C$_3$ corresponds to consonant /T/, GS corresponds to glottal stop substitution, PA corresponds to palatalized articulation, PSNAE corresponds to phoneme specific nasal air emission, and velar corresponds to velar substitutions.}}
	\label{T_7}
	\scalebox{0.74}{
		\begin{tabular}{ccccccccc}
			\hline \hline
			\multirow{2}{*}{Words} &  &  \multirow{2}{*}{CLP$_{\text{original}}$} &  & \multicolumn{5}{c}{CLP$_{\text{modified}}$}          \\ \cline{5-9} 
			&  &                                                                    &  & Obstruent &  & Vowel &  & Obstruent + vowel \\ \hline
			/FVFV/ (GS)              &  &1.00$\pm$0.09             &  & 1.5$\pm$0.07     &  &2.10$\pm$0.15             & &3.10$\pm$0.50   \\
			/FVFV/ (PA)              &  &1.10$\pm$0.15             &  &1.54$\pm$0.55      &  & 2.42$\pm$0.26             & &3.92$\pm$0.50         \\
			/FVFV/ (PSNAE)           &  &1.50$\pm$0.40              &  &1.90$\pm$0.60      &  & 2.50$\pm$0.20             & &3.80$\pm$0.20  \\ \hline
			/C$_1$VC$_1$V/ (GS)      &  &1.58$\pm$1.17               &  &1.88$\pm$0.75      &  & 2.85$\pm$0.46            & &3.72$\pm$0.45  \\ \hline
			/C$_2$VC$_2$V/ (GS)      &  &1.12$\pm$1.11              &  &1.53$\pm$0.51      &  & 2.27$\pm$0.47            & &3.57$\pm$0.99  \\
			/C$_2$VC$_2$V/ (Velar)   &  &1.08$\pm$1.19              &  & 1.97$\pm$0.99      &  & 2.92$\pm$0.88            & &3.82$\pm$0.55  \\
			/C$_2$VC$_2$V/ (PA)      &  &1.94$\pm$0.98              &  & 2.02$\pm$0.58     &  & 2.80$\pm$0.76            & &3.71$\pm$0.87  \\ \hline
			/C$_3$VC$_3$V/ (GS)      &  &1.20$\pm$1.01              &  & 1.54$\pm$0.74     &  & 2.10$\pm$0.92            & &2.99$\pm$0.22  \\
			/C$_3$VC$_3$V/ (Velar)   &  &1.15$\pm$0.83              &  & 1.72$\pm$1.20     &  & 2.30$\pm$0.57            & &3.05$\pm$1.09  \\
			/C$_3$VC$_3$V/ (PA)      &  &1.53$\pm$1.25              &  & 1.72$\pm$0.58     &  & 2.70$\pm$0.89             & &3.59$\pm$1.12  \\  \hline\hline
		\end{tabular}
	}
\end{table}
A total of 10 naive listeners have participated in the study and the speech samples were randomly numbered to avoid any bias towards the original or modified speech. The listeners bear the knowledge of speech science. Each of the listener have evaluated 120 speech samples (3 vowel contexts $\times$ 3 fricative /s/ errors $\times$ 4 variations = 36, 3 vowel contexts $\times$ 1 stop /k/ error $\times$ 4 variations = 12, 3 vowel contexts $\times$ 3 stop /t/ errors $\times$ 4 variations = 36, 3 vowel contexts $\times$ 3 stop /T/ errors $\times$ 4 variations = 36). The MOS  values are derived by averaging all the values across each vowel context of the specific word from all the listeners. The averaged MOS values shown in Table~\ref{T_7} indicate that the modification of all phonemes provide significant improvement compared to the original and standalone modified CLP speech.

This study  has been performed on speech data collected in clinical settings from children having mild to moderate CLP speech disorder. The scope of the work is limited to studying specific CLP speech distortions in isolated phonemes in the context of /CVCV/ words with identical /CV/ pairs.  The study primarily focuses on enhancing a few select obstruents and vowels. Later, it is extended to some nonsensical  /CVCV/ words that can be formed by combining the studied obstruents and vowels. The proposed system is developed based on certain assumptions in a controlled environment. Therefore, realization of the proposed techniques in real-time deserves future explorations.

\section{Conclusion}
The word-level intelligibility is attempted by combining the specific phoneme modifications discussed in our previous works. A comparison study is also done to observe whether the transformation method exploited in our earlier works can improve the entire word intelligibility. When the transformation method is used to modify the word as a whole, it is observed that the speech sounds muffled. Hence, phoneme specific modifications are exploited to observe its impact in word intelligibility. From the evaluation results, it is observed that, a significant improvement in the word-level intelligibility can be achieved when all phonemes in the words are modified independently. 

\section{Acknowledgement}
The authors would like to thank Dr. M. Pushpavathi and Dr. Ajish Abraham, AIISH Mysore, for providing insights about CLP speech disorder. The authors would also like to acknowledge the research scholars of IIT Guwahati for their participation in the perceptual test. This work is in part supported by a project entitled “NASOSPEECH: Development of Diagnostic system for Severity Assessment of the Disordered Speech” funded by the Department of Biotechnology (DBT), Govt. of India.

\bibliographystyle{IEEEtran}

\bibliography{enh_ch_8_new1}

\begin{thebibliography}{10}
\providecommand{\url}[1]{#1}
\csname url@samestyle\endcsname
\providecommand{\newblock}{\relax}
\providecommand{\bibinfo}[2]{#2}
\providecommand{\BIBentrySTDinterwordspacing}{\spaceskip=0pt\relax}
\providecommand{\BIBentryALTinterwordstretchfactor}{4}
\providecommand{\BIBentryALTinterwordspacing}{\spaceskip=\fontdimen2\font plus
\BIBentryALTinterwordstretchfactor\fontdimen3\font minus
  \fontdimen4\font\relax}
\providecommand{\BIBforeignlanguage}[2]{{%
\expandafter\ifx\csname l@#1\endcsname\relax
\typeout{** WARNING: IEEEtran.bst: No hyphenation pattern has been}%
\typeout{** loaded for the language `#1'. Using the pattern for}%
\typeout{** the default language instead.}%
\else
\language=\csname l@#1\endcsname
\fi
#2}}
\providecommand{\BIBdecl}{\relax}
\BIBdecl

\bibitem{kain2007improving}
A.~B. Kain, J.-P. Hosom, X.~Niu, J.~P. van Santen, M.~Fried-Oken, and
  J.~Staehely, ``Improving the intelligibility of dysarthric speech,''
  \emph{Speech Communication}, vol.~49, no.~9, pp. 743--759, 2007.

\bibitem{rudzicz2013adjusting}
F.~Rudzicz, ``Adjusting dysarthric speech signals to be more intelligible,''
  \emph{Computer Speech \& Language}, vol.~27, no.~6, pp. 1163--1177, 2013.

\bibitem{nakamura2012speaking}
K.~Nakamura, T.~Toda, H.~Saruwatari, and K.~Shikano, ``Speaking-aid systems
  using {GMM}-based voice conversion for electrolaryngeal speech,''
  \emph{Speech Communication}, vol.~54, no.~1, pp. 134--146, 2012.

\bibitem{whitehill2002assessing}
T.~L. Whitehill, ``Assessing intelligibility in speakers with cleft palate: a
  critical review of the literature,'' \emph{The Cleft Palate-Craniofacial
  Journal}, vol.~39, no.~1, pp. 50--58, 2002.

\bibitem{lai2019multi}
Y.-H. Lai and W.-Z. Zheng, ``Multi-objective learning based speech enhancement
  method to increase speech quality and intelligibility for hearing aid device
  users,'' \emph{Biomedical Signal Processing and Control}, vol.~48, pp.
  35--45, 2019.

\bibitem{biadsy2019parrotron}
F.~Biadsy, R.~J. Weiss, P.~J. Moreno, D.~Kanevsky, and Y.~Jia, ``Parrotron: An
  end-to-end speech-to-speech conversion model and its applications to
  hearing-impaired speech and speech separation,'' in \emph{Proceedings of
  Interspeech}, 2019, pp. 4115--4119.

\bibitem{fu2017joint}
S.-W. Fu, P.-C. Li, Y.-H. Lai, C.-C. Yang, L.-C. Hsieh, and Y.~Tsao, ``Joint
  dictionary learning-based non-negative matrix factorization for voice
  conversion to improve speech intelligibility after oral surgery,'' \emph{IEEE
  Transactions on Biomedical Engineering}, vol.~64, no.~11, pp. 2584--2594,
  2017.

\bibitem{aihara2013individuality}
R.~Aihara, R.~Takashima, T.~Takiguchi, and Y.~Ariki, ``Individuality-preserving
  voice conversion for articulation disorders based on non-negative matrix
  factorization,'' in \emph{Proceedings of IEEE International Conference on
  Acoustics, Speech and Signal Processing (ICASSP)}, 2013, pp. 8037--8040.

\bibitem{xiao2019reconstruction}
K.~Xiao, S.~Wang, M.~Wan, and L.~Wu, ``Reconstruction of mandarin
  electrolaryngeal fricatives with hybrid noise source,'' \emph{IEEE/ACM
  Transactions on Audio, Speech and Language Processing (TASLP)}, vol.~27,
  no.~2, pp. 383--391, 2019.

\bibitem{bi1997application}
N.~Bi and Y.~Qi, ``Application of speech conversion to alaryngeal speech
  enhancement,'' \emph{IEEE Transactions on Speech and Audio Processing},
  vol.~5, no.~2, pp. 97--105, 1997.

\bibitem{kummer2013cleft}
A.~W. Kummer, \emph{Cleft Palate \& Craniofacial Anomalies: Effects on Speech
  and Resonance}.\hskip 1em plus 0.5em minus 0.4em\relax Nelson Education,
  2013.

\bibitem{peterson2001cleft}
S.~J. Peterson-Falzone, M.~A. Hardin-Jones, and M.~P. Karnell, \emph{Cleft
  Palate Speech}.\hskip 1em plus 0.5em minus 0.4em\relax Mosby St. Louis, 2001.

\bibitem{henningsson2008universal}
G.~Henningsson, D.~P. Kuehn, D.~Sell, T.~Sweeney, J.~E. Trost-Cardamone, and
  T.~L. Whitehill, ``Universal parameters for reporting speech outcomes in
  individuals with cleft palate,'' \emph{The Cleft Palate-Craniofacial
  Journal}, vol.~45, no.~1, pp. 1--17, 2008.

\bibitem{scipioni2009intelligibility}
M.~Scipioni, M.~Gerosa, D.~Giuliani, E.~N{\"o}th, and A.~Maier,
  ``Intelligibility assessment in children with cleft lip and palate in
  {I}talian and {G}erman,'' in \emph{Tenth Annual Conference of the
  International Speech Communication Association}, 2009.

\bibitem{maier2006intelligibility}
A.~Maier, C.~Hacker, E.~Noth, E.~Nkenke, T.~Haderlein, F.~Rosanowski, and
  M.~Schuster, ``Intelligibility of children with cleft lip and palate:
  {E}valuation by speech recognition techniques,'' in \emph{18th International
  Conference on Pattern Recognition (ICPR)}, vol.~4.\hskip 1em plus 0.5em minus
  0.4em\relax IEEE, 2006, pp. 274--277.

\bibitem{grunwell2001speech}
P.~Grunwell and D.~Sell, ``Speech and cleft palate/velopharyngeal anomalies,''
  \emph{Management of Cleft Lip and Palate. London: Whurr}, 2001.

\bibitem{vikram2016spectral}
C.~Vikram, N.~Adiga, and S.~M. Prasanna, ``Spectral enhancement of cleft lip
  and palate speech.'' in \emph{Proceedings of Interspeech}, 2016, pp.
  117--121.

\bibitem{sudro2021modification}
P.~N. Sudro and S.~M. Prasanna, ``Modification of misarticulated fricative /s/
  in cleft lip and palate speech,'' \emph{Biomedical Signal Processing and
  Control}, vol.~67, p. 102088, 2021.

\bibitem{sudro2021event}
P.~N. Sudro, C.~Vikram, and S.~M. Prasanna, ``Event-based transformation of
  misarticulated stops in cleft lip and palate speech,'' \emph{Circuits,
  Systems, and Signal Processing}, vol.~40, no.~8, pp. 4064--4088, 2021.

\bibitem{sudro2020enhancement}
P.~N. Sudro and S.~M. Prasanna, ``Enhancement of cleft palate speech using
  temporal and spectral processing,'' \emph{Speech Communication}, vol. 123,
  pp. 70--82, 2020.

\bibitem{slm_recording}
\emph{Bruel and Kjaer}, 1942, https://www.bksv.com/en.

\bibitem{singh2016structure}
K.~Singh and N.~Tiwari, ``The structure of {Hindi} stop consonants,'' \emph{The
  Journal of the Acoustical Society of America}, vol. 140, no.~5, pp.
  3633--3642, 2016.

\bibitem{stylianou1998continuous}
Y.~Stylianou, O.~Capp{\'e}, and E.~Moulines, ``Continuous probabilistic
  transform for voice conversion,'' \emph{IEEE Transactions on Speech and Audio
  Processing}, vol.~6, no.~2, pp. 131--142, 1998.

\bibitem{krishnamoorthy2011enhancement}
P.~Krishnamoorthy and S.~R.~M. Prasanna, ``Enhancement of noisy speech by
  temporal and spectral processing,'' \emph{Speech Communication}, vol.~53,
  no.~2, pp. 154--174, 2011.

\bibitem{stevens2000acoustic}
K.~N. Stevens, \emph{Acoustic phonetics}.\hskip 1em plus 0.5em minus
  0.4em\relax MIT press, 2000, vol.~30.

\bibitem{narayanan2000noise}
S.~Narayanan and A.~Alwan, ``Noise source models for fricative consonants,''
  \emph{IEEE Transactions on Speech and Audio processing}, vol.~8, no.~3, pp.
  328--344, 2000.

\bibitem{shadle1985acoustics}
C.~H. Shadle, ``The acoustics of fricative consonants,'' 1985.

\bibitem{shadle1996quantifying}
C.~H. Shadle and S.~J. Mair, ``Quantifying spectral characteristics of
  fricatives,'' in \emph{Proceedings of 4th International Conference on Spoken
  Language Processing (ICSLP)}, vol.~3, 1996, pp. 1521--1524.

\bibitem{prathosh2014estimation}
A.~Prathosh, A.~Ramakrishnan, and T.~Ananthapadmanabha, ``Estimation of
  voice-onset time in continuous speech using temporal measures,'' \emph{The
  Journal of the Acoustical Society of America}, vol. 136, no.~2, pp.
  EL122--EL128, 2014.

\bibitem{delattre1955acoustic}
P.~C. Delattre, A.~M. Liberman, and F.~S. Cooper, ``Acoustic loci and
  transitional cues for consonants,'' \emph{The Journal of the Acoustical
  Society of America}, vol.~27, no.~4, pp. 769--773, 1955.

\bibitem{janbakhshi2019pathological}
P.~Janbakhshi, I.~Kodrasi, and H.~Bourlard, ``Pathological speech
  intelligibility assessment based on the short-time objective intelligibility
  measure,'' in \emph{Proceedings of IEEE International Conference on
  Acoustics, Speech and Signal Processing (ICASSP)}, 2019, pp. 6405--6409.

\bibitem{povey2011kaldi}
D.~Povey, A.~Ghoshal, G.~Boulianne, L.~Burget, O.~Glembek, N.~Goel,
  M.~Hannemann, P.~Motlicek, Y.~Qian, P.~Schwarz \emph{et~al.}, ``The kaldi
  speech recognition toolkit,'' in \emph{IEEE 2011 workshop on automatic speech
  recognition and understanding}, no. CONF.\hskip 1em plus 0.5em minus
  0.4em\relax IEEE Signal Processing Society, 2011.

\end{thebibliography}

\end{document}